# SAGE-Flow: A Decoupled Framework for Geometry Alignment and Stateless Real-Time Multi-Camera 3D Reconstruction

Chentian Sun

***Abstract*—Real-time multi-camera 3D reconstruction remains challenging due to the strong dependency among extrinsic calibration, multi-view fusion, and global optimization, which limits reconstruction stability and scalability.**

**This paper presents SAGE-Flow, a decoupled framework consisting of geometry-aligned multi-view calibration (GMAC) and stateless adaptive geometric representation (SAGE). GMAC estimates camera extrinsics from geometric constraints without calibration targets, dense images, or bundle adjustment. SAGE constructs a compact geometric representation by selecting reliable multi-view observations under a bounded geometric budget, achieving linear time and memory complexity.**

**Experiments show SAGE-Flow achieves precise camera calibration, low 3D reconstruction cost, good scalability, and can generate high-quality point clouds under limited throughput.**

## I. INTRODUCTION

Multi-camera real-time 3D reconstruction (MRR) is a key technology for immersive multimedia applications such as volumetric video, telepresence, VR/AR [1]–[5]. Multi-camera setups are increasingly accessible, but real-time 3D output from multi-view video is still challenging [6]–[8].

Existing MRR systems have four key drawbacks. 1. Unstable extrinsic calibration. Cross-view alignment demands precise camera poses. Traditional calibration is vulnerable to vibration, thermal shift and scene variation, and relies on expensive bundle adjustment [9]–[12]. 2. Unreliable geometry from redundant multi-view data [13]–[15]. Occlusion, reflection and sensor noise corrupt depth readings [16]–[18]. Prior methods cost heavy computation and fail easily under pose errors. Direct geometry fusion neglects observation trust and compact representation, causing redundant data transfer and shaky reconstruction. 3. Poor scalability. TSDF, SLAM and neural reconstruction need global states or iterative optimization, leading to super-linear computation cost [19]–[22]. 4. Bottlenecked data capture. Conventional schemes reduce data size only after capture. This wastes capture bandwidth and restricts multi-camera setups [30]–[34].

To address this issue, a decoupled framework for geometry alignment and stateless real-time multi-view 3D reconstruction, termed SAGE-Flow, is proposed. The framework consists of geometry-aligned multi-view calibration (GMAC) and stateless adaptive geometric representation (SAGE).

GMAC leverages cross-view reprojection and cycle consistency to derive metric camera extrinsics from feed-forward multi-view backbones (MRB), a family of modern geometric large models exemplified by VGGT [9] and MapAnything [10]. It delivers target-free RGB-D calibration and avoids bundle adjustment entirely. SAGE adopts a stateless design, scoring observation quality via confidence and cross-view consistency to produce compact geometric representation with linear complexity. GMAC delivers accurate camera poses, and SAGE refines aligned observations into concise scene geometry while mitigating minor alignment errors for high-performance real-time reconstruction.

As shown in Fig. 1, SAGE-Flow focuses on point cloud acquisition and generation for telepresence, reducing calibration costs and enabling high-frame-rate, high-quality point clouds under limited throughput.

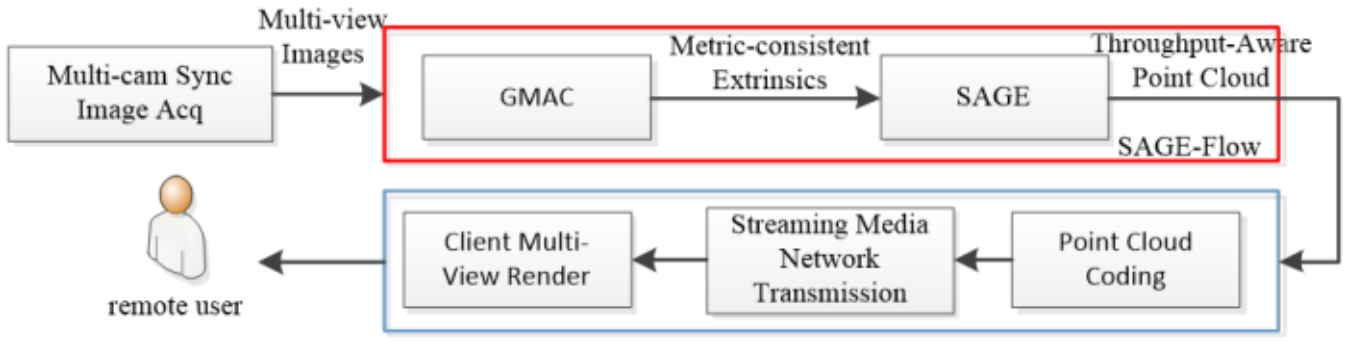

**Fig. 1** Utilization of SAGE-Flow in Remote Multimedia Interaction

The contributions of this paper are as follows:

1. A decoupled MRR framework that separates calibration and fusion to reduce coupling and improve scalability.

2. GMAC, a metric-scale lifting calibration method that transforms relative-scale pose predictions into physically grounded extrinsics using reprojection and cycle-consistency constraints, without calibration targets, global bundle adjustment, or additional training.

3. SAGE, a stateless confidence-driven geometric representation method with linear complexity designed for real-time reconstruction under a limited throughput budget.

4. SAGE-Flow, an integrated pipeline that demonstrates the synergy between calibration and reconstruction for practical multi-camera 3D reconstruction, optimized for low-latency remote telepresence and VR/AR immersive applications.

## II. RELATED WORK

Multi-camera 3D reconstruction involves three aspects: extrinsic calibration, multi-view fusion, and system scalability.

### *A. Multi-Camera Extrinsic Calibration*

Traditional calibration methods use checkerboards to establish 2D–3D correspondences [1]. Although accurate in controlled environments, they require fixed camera layouts and manual recalibration after any mechanical change [2], [3].

Online approaches based on SLAM/VIO, such as ORB-SLAM3 [4], jointly optimize poses and maps through bundle adjustment, but their computational cost increases rapidly with camera count [5]. Learning-based methods, including VGGSfM [6], DFSfM [7], PoseDiffusion [8], VGGT [9], and MapAnything [10], provide strong pose estimates from images alone. However, they are primarily designed for offline reconstruction and cannot reliably adapt to unseen scene scales.

Target-free extrinsic refinement methods remain limited [11]–[13]. GMAC addresses this problem by combining modern MRB with a geometry-based optimization strategy that supports large camera arrays.

### *B. Multi-View 3D Fusion and Point Cloud Reconstruction*

Volumetric methods such as KinectFusion [14] reconstruct dense geometry using TSDF integration, but memory and computation increase with scene volume and performance degrades in dynamic environments [15], [16]. Surfel- and mesh-based systems, including SurfelMeshing [17] and ElasticFusion [18], generate smoother surfaces but rely on persistent global maps and are prone to accumulated registration errors [19], [20].

R3D3 [21] and VR2Gather [35] extends SLAM to multi-camera dense reconstruction but still suffers from dynamic-scene degradation and growing computational cost. Neural approaches, including NeRF [22] and 3D Gaussian Splatting [23], achieve high reconstruction quality but require calibrated cameras and expensive offline optimization.

Uncertainty-aware fusion methods improve robustness through confidence modeling [24], [25], yet most still depend on global TSDF optimization [26]–[29]. In contrast, SAGE performs stateless geometric representation from multi-view observations without maintaining a global scene representation.

### *C. Scalability in Multi-Camera Systems*

Existing telepresence and volumetric video frameworks boost scalability via geometry compression, coding [30], [31], and adaptive streaming [32]–[35] to cut transmission costs. While these methods greatly improve throughput efficiency, they only work after point cloud capture and rely on fully reconstructed dense geometry.

SAGE-Flow takes a different approach by accounting for later representation and transmission during multi-camera point cloud generation. Its stateless geometric geometry eliminates redundant data at capture stage and generates compact geometric primitives before compression or streaming. This lowers reconstruction complexity and raw data overhead for all downstream processing.

### *D. Summary*

Existing methods address only part of the problem. Offline approaches provide high accuracy but lack real-time capability [29]–[32]. Online SLAM systems support real-time operation but suffer from drift and limited scalability, while learning-based methods lack online refinement mechanisms. SAGE-Flow combines calibration and linear-complexity reconstruction within a unified real-time framework.

## III. PROPOSED METHOD

### *A. Framework Overview*

Considering a synchronized multi-camera system composed of $N_{cam}$ cameras, at time step $t$, the system receives $\{I_i^t, D_i^t\}_{i=1}^{N_{cam}}$, where $I_i^t \in \mathbb{R}^{H\times W\times 3}$ and $D_i^t \in \mathbb{R}^{H\times W}$ denote the RGB image and depth map of camera $i$, respectively (Depth comes from RGB-D depth maps or RGB depth estimation. To keep things uniform, RGB-D cameras are used as standard setup in this paper). Camera intrinsics $K_i$ are assumed known, while extrinsics $\{\boldsymbol{T}_i\}_{i=1}^{N_{cam}}$ are unknown:

$$\boldsymbol{T}_i^t = [\boldsymbol{R}_i^t | \boldsymbol{t}_i^t] \in \boldsymbol{SE}(3) \tag{1}$$

Three correlated technical tasks are defined for this system.

Task 1 (GMAC): Given $\{I_i^t, D_i^t\}_{i=1}^{N_{cam}}$, estimate extrinsics $\boldsymbol{T}_i^t$ with a little manual operation, compatible with any MRB.

Task 2 (SAGE): Given $\{I_i^t, D_i^t\}_{i=1}^{N_{cam}}$ and extrinsics $\boldsymbol{T}_i^t$ from any source, generate a compact reconstructed point cloud $P^t$ through SAGE with $O(N_{cam}HW)$ per-frame complexity.

Task 3 (SAGE-Flow): Compose GMAC and SAGE into a unified pipeline, and evaluate the synergistic gains.

Extrinsic calibration and point cloud reconstruction differ in optimization goals and running frequency. Calibration estimates slowly changing camera poses, while reconstruction processes continuous streaming measurements. Joint optimization creates redundant coupling between these two dissimilar tasks complicating deployment. SAGE-Flow uses a decoupled design enabling independent optimization and execution for each component.

GMAC builds on MRB. MRB extract far stronger cross-view geometric priors than conventional extrinsic prediction pipelines, yet they are not built for pose estimation and cannot directly output metric-scale extrinsics for multi-camera setups. GMAC leverages lightweight geometric optimization to recover physically consistent metric poses, adapting MRBs to multi-camera extrinsic calibration. Once accurate extrinsics are available, SAGE employs a stateless reconstruction scheme that only processes current-frame inputs. This eliminates drift from accumulated historical errors, caps memory consumption, and keeps computation linear. The two modular components work across diverse environments, forming an efficient, scalable real-time reconstruction pipeline.

Considering the stateless nature of the proposed method without historical state accumulation, we omit the time-step $t$ in the following derivations for notation simplicity, and all fixed hyperparameters follow the empirically tested settings.

### *B. GMAC: Geometry-Aware Multi-View Extrinsic Calibration*

#### *1) Motivation of GMAC*

Recently proposed MRB methods are able to infer globally consistent camera geometry from only few images without calibration targets or dense correspondences. However, their predictions suffer from global scale ambiguity: the translation directions and ratios are preserved, but the absolute scale lacks physical constraints and cannot be directly used for metric extrinsic estimation in RGB-D systems.

GMAC bridges this gap by using MRB to obtain relative poses and introducing a graph-consistent metric lifting strategy to recover physically scaled extrinsics. Instead of pursuing laboratory-level accuracy, GMAC targets assisted calibration for large camera arrays, reducing manual effort.

*2) MRB Feature Extraction and Extrinsic Initialization*

Let $\phi$ denote the MRB network, to focus on the scale-lifting task, task-irrelevant components are removed while preserving the core modules for cross-view geometric reasoning:

1. Retain: Multi-scale feature extraction, cross-view geometric attention, global latent feature fusion — the modules responsible for cross-view geometric reasoning and can output globally coupled geometric representations.

2. Remove: Dense 3D rendering decoders, pixel-level view synthesis branches, and point cloud generation modules irrelevant to calibration.

With the backbone frozen and only lightweight inference modules added, the encoder produces the scale-ambiguous initial extrinsic $\{\boldsymbol{T}_i^0\} = \{[\boldsymbol{R}_i|\boldsymbol{t}_i^0]\}$.

*3) Graph-Consistent Metric Lifting*

- Scale Ambiguity and Graph Coupling

The MRB enforces that inter-camera distance ratios are approximately correct across the graph but leaves absolute magnitudes unanchored. An unknown scale factor $s^* = \mathbb{R}^+$, from global coupling, maps encoder outputs to metric poses:

$$\boldsymbol{t}_i^{metric} = s^* \cdot \boldsymbol{t}_i^0, \forall i \in \mathcal{V} \tag{2}$$

- Graph Coupled Scale Consistency Objective

For each overlapping pair $(i,j) \in \mathcal{E}$, a match set $\mathcal{U}_{ij}$ is built via depth-guided reprojection consistency (RC). For pixel $\boldsymbol{u} \in \mathcal{U}_{ij}$, camera-$i$ 3D point:

$$\boldsymbol{X}_i(\boldsymbol{u}) = \pi_i^{-1}(\boldsymbol{u}, D_i(\boldsymbol{u}), K_i) \tag{3}$$

Under global scale $s$, the relative transformation $\boldsymbol{T}_{ji} = [\boldsymbol{R}_{ji}|\boldsymbol{t}_{ji}]$ maps $i$-frame points into camera $j$, yielding a depth component linear in $s$ ($Z(\cdot)$: predicted depth in camera):

$$Z(\boldsymbol{X}_{ij}(\boldsymbol{u}, s)) = s(\boldsymbol{R}_{ji}\boldsymbol{X}_i(\boldsymbol{u}))_z + (\boldsymbol{t}_{ji})_z \tag{4}$$

The scale consistency objective aggregates depth residuals across all graph edges:

$$L_{scale}(s) = \sum_{(i,j)\in\mathcal{E}} \sum_{\boldsymbol{u}\in\mathcal{U}_{ij}} \omega_{ij}^{GMAC}(\boldsymbol{u}) \left\| Z\left(\boldsymbol{X}_{ij}(\boldsymbol{u}, s)\right) - D_j\left(\pi(\boldsymbol{X}_{ij}(\boldsymbol{u}, s))\right) \right\|_2^2 \tag{5}$$

$\omega_{ij}^{GMAC}(\boldsymbol{u})$ is the GMAC matching weight for $\boldsymbol{u}$.

The per-correspondence weight encodes graph-level observability and pixel-level depth validity:

$$\omega_{ij}^{GMAC}(\boldsymbol{u}) = o_{ij} \cdot cos\langle \boldsymbol{F}_i, \boldsymbol{F}_j \rangle \cdot \omega_i^{depth}(\boldsymbol{u}) \cdot \omega_j^{depth}(\pi(\boldsymbol{X}_{ij}(\boldsymbol{u}))) \tag{6}$$

where $o_{ij} \in [0.2,1]$ gates edges with insufficient overlap; $cos\langle \boldsymbol{F}_i, \boldsymbol{F}_j \rangle$ is the cosine similarity between latent feature vectors serving as a proxy for structural compatibility; and $\omega_i^{depth}(\boldsymbol{u})$ is the sensor validity mask. A correspondence contributes strongly only when all three conditions are simultaneously satisfied, making the estimator robust to partial overlap, latent mismatch, and sensor noise within a single unified weighting scheme.

- Closed Form Metric Lifting Solution

Under the fixed-correspondence approximation — lookup coordinates are evaluated at current correspondence geometry and not updated as $s$ varies —$L_{scale}$ reduces to a quadratic function of $s$. Let

$$a_{ij}(\boldsymbol{u}) = (\boldsymbol{R}_{ji}\boldsymbol{X}_i(\boldsymbol{u}))_z \tag{7}$$

$$b_{ij}(\boldsymbol{u}) = (\boldsymbol{t}_{ji})_z \tag{8}$$

$$d_{ij}(\boldsymbol{u}) = D_j(\pi(\boldsymbol{T}_{ji}\boldsymbol{X}_i(\boldsymbol{u}))) \tag{9}$$

Setting $\frac{\mathrm{d}L_{scale}}{ds} = 0$ yields the unique closed-form minimizer for the global metric scale $s^*$:

$$s^* = \frac{\sum_{(i,j)\in\mathcal{E}} \sum_{\boldsymbol{u}\in\mathcal{U}_{ij}} \omega_{ij}^{GMAC}(\boldsymbol{u})(d_{ij}(\boldsymbol{u}) - b_{ij}(\boldsymbol{u}))\cdot a_{ij}(\boldsymbol{u})}{\sum_{(i,j)\in\mathcal{E}} \sum_{\boldsymbol{u}\in\mathcal{U}_{ij}} \omega_{ij}^{GMAC}(\boldsymbol{u})\cdot a_{ij}^2(\boldsymbol{u})} \tag{10}$$

This estimator requires no iterative optimization, making per-frame scale recovery feasible within the inference budget of a real-time system. The estimator is uniquely defined whenever at least one valid correspondence exists with nonzero rotation-projected depth, a condition satisfied for any non-degenerate scene under graph connectivity.

- Cycle-Consistent Graph Validation

Accumulated encoder errors or outlier edges can introduce local scale inconsistencies that may propagate around graph cycles and degrade global metric coherence. For triplet $(i,j,k)$ with $(i,j)$, $(j,k)$, $(k,i) \in \mathcal{E}$, the cycle transformation is $\boldsymbol{T}_{ijk}^{cycle} = \boldsymbol{T}_{ik} \cdot \boldsymbol{T}_{kj} \cdot \boldsymbol{T}_{ji}$. The normalized cycle residual over sampled 3D reference points $\{\boldsymbol{X}^{(q)}\}_{q=1}^{N_{samp}}$ is:

$$\varepsilon_{cycle} = \frac{1}{N_{samp}\cdot|\Gamma|} \sum_{q=1}^{N_{samp}} \sum_{(i,j,k)\in\Gamma} \left\| \boldsymbol{T}_{ijk}^{cycle}\boldsymbol{X}^{(q)} - \boldsymbol{X}^{(q)} \right\|_2^2 \tag{11}$$

where Γ is the set of valid triplets in $G$. The recovered metric configuration is accepted when $\varepsilon_{cycle} < \delta_{cycle} = 0.01$. When the criterion is not satisfied, edges contributing most to the cycle residual are excluded and $s^*$ is re-estimated on the pruned graph. This outlier rejection typically converges within a small number of iterations, as cycle-inconsistent edges constitute a small minority of the full edge set.

- Metric-Consistent Extrinsics

The final metric-consistent extrinsics are:

$$\boldsymbol{T}_i^{metric} = [\boldsymbol{R}_i | s^* \cdot \boldsymbol{t}_i^0] \tag{12}$$

Applying $s^*$ uniformly preserves all inter-camera directional relationships while establishing absolute scale from sensor depth. The resulting metric-consistent extrinsics are directly usable in any downstream RGB-D reconstruction task without further scaling, and serve as the input for the SAGE module.

*C. SAGE: Stateless Adaptive Geometric Representation*

SAGE is a stateless geometric representation module designed for real-time RGB-D camera arrays. It constructs a compact geometric representation directly from synchronized multi-view observations without relying on historical states. Given calibrated camera poses $\{\boldsymbol{T}_i\}_{i=1}^{N_{cam}}$ and synchronized measurements captured from $N_{cam}$ cameras, the system generates an initial candidate geometric point set: $\mathcal{P} = \{\boldsymbol{X}_m\}_{m=1}^{M_{total}}$, where $M_{total}$ refers to the total count of valid depth observation points.

SAGE formulates multi-view reconstruction as a reliability-aware geometric representation problem under a bounded geometric budget, that sustains high reconstruction quality under a limited representation budget.

*1) Stateless Geometric Stream Processing*

In SAGE, the reconstruction output at timestamp $t$ is determined solely by the frame-specific geometric observation set $\mathcal{P}_t$. Historical geometric states are completely decoupled from the reconstruction of the current frame.

This framework is inherently suitable for dynamic scenarios and enables large-scale real-time multi-camera reconstruction tasks to be feasible. On the other hand, large-scale real-time multi-camera reconstruction brings heavy rendering overhead and throughput overhead—novel challenges unaddressed by existing reconstruction methods designed for small-scale scenes or offline applications.

For an array with $N_{cam}$ cameras running at frame rate $F$ and per-frame resolution $H \times W$, the number of valid depth points scales as $M_{total} = \mathrm{O}(N_{cam} HWF)$, a challenge rarely addressed by traditional reconstruction pipelines.

To resolve this bottleneck, SAGE enforces an explicit geometric budget constraint: $|\mathcal{P}_f| \leq B_{max}$, where $\mathcal{P}_f$ denotes the final reconstructed output point cloud. Under this constraint, reconstruction is transformed into a constrained geometric quality maximization problem:

$$\mathcal{P}_f^* = \arg\max_{\mathcal{P}_f \subseteq P} Q(\mathcal{P}_f), \text{s.t.} |\mathcal{P}_f| \leq B_{max} \tag{13}$$

The objective function $Q(\cdot)$ aims to preserve the most informative geometric representation in a fixed storage budget.

*2) Uncertainty-Aware Geometric Quality Modeling*

Since the stateless pipeline contains no temporal refinement steps, geometric quality must be directly estimated from raw single-frame observations.

For any candidate 3D point $\boldsymbol{X} \in \mathcal{P}$, SAGE computes two complementary quality metrics: single-frame measurement reliability (MR), multi-view geometric consistency (CVGC).

Normalized depth gradient and local depth variance are utilized to quantify local depth instability;

$$g(\boldsymbol{u}) = \frac{\|\nabla D(\boldsymbol{u})\|}{D(\boldsymbol{u})+\epsilon}, v(\boldsymbol{x}) = \frac{Var\left(D_{\mathcal{N}_{pix}(\boldsymbol{u})}\right)}{(D(\boldsymbol{u})+\epsilon)^2} \tag{14}$$

where $\mathcal{N}_{pix}(\boldsymbol{u})$ is the pixel neighborhood centered at $\boldsymbol{u}$. The single-view measurement confidence is defined as:

$$c_m(\boldsymbol{u}) = \exp\left(-\alpha g(\boldsymbol{u}) - \beta v(\boldsymbol{u})\right) \tag{15}$$

where $\alpha = 1.2, \beta = 0.8$ . To further discard geometrically inconsistent observations, each 3D candidate point $\boldsymbol{X}$ is reprojected to the frustum of adjacent cameras in neighbor set $\mathcal{N}_{cam}(i)$, and cross-view consistency scores are computed based on normalized depth reprojection errors:

$$c_v(\boldsymbol{X}) = \frac{1}{|\mathcal{N}_{cam}(i)|} \sum_{j \in \mathcal{N}_{cam}(i)} \exp\left(-\tau e_{ij}\right) \tag{16}$$

where $\tau = 5$, $e_{ij}$ represents the normalized deviation between reprojected depth and raw observed depth from camera $j$.

The final unified confidence score is calculated via harmonic mean to balance single-view and multi-view quality signals:

$$\mathrm{r}(\boldsymbol{X}) = \frac{2c_m(\boldsymbol{X})c_v(\boldsymbol{X})}{c_m(\boldsymbol{X})+c_v(\boldsymbol{X})} \tag{17}$$

The harmonic mean naturally suppresses two categories of low-quality points: samples suffering severe measurement noise, and points with inconsistent cross-view matching results.

*3) Budget-Aware Geometric Representation Objective*

A naive geometric budget constraint alone cannot produce uniform spatial coverage. If points are filtered purely by raw confidence scores, sampling density will concentrate heavily in highly textured regions, leading to incomplete scene coverage.

SAGE formulates geometric representation as a budget-aware objective balancing quality and spatial uniformity. The total utility of the selected point subset is defined as the sum of confidence values for retained 3D points:

$$U(\mathcal{P}_f) = \sum_{\boldsymbol{X}_m \in \mathcal{P}_f} r(\boldsymbol{X}_m) \tag{18}$$

This term prioritizes observations with higher measurement reliability. A spatial redundancy penalty term is introduced to mitigate over-sampling in spatially overlapping regions:

$$\Omega(\mathcal{P}_f) = \sum_{m<n} \exp\left(-\frac{\|\boldsymbol{X}_m - \boldsymbol{X}_n\|^2}{\sigma_m^2 + \sigma_n^2}\right) \tag{19}$$

where $\sigma_m$ denotes the local geometric uncertainty of the $m$-th candidate 3D point $\boldsymbol{X}_m$, $\sigma_m = 5cm$ in this paper.

The geometric selection objective combines total utility and redundancy regularization:

$$\mathrm{Q}(\mathcal{P}_f) = U(\mathcal{P}_f) - \lambda\Omega(\mathcal{P}_f) \tag{20}$$

where $\lambda = 0.1$. The above objective defines the desired point subset selection criterion. To enable real-time reconstruction, SAGE decomposes Eq. (20) into regional optimization, where each spatial cell maximizes local utility under the budget constraint.

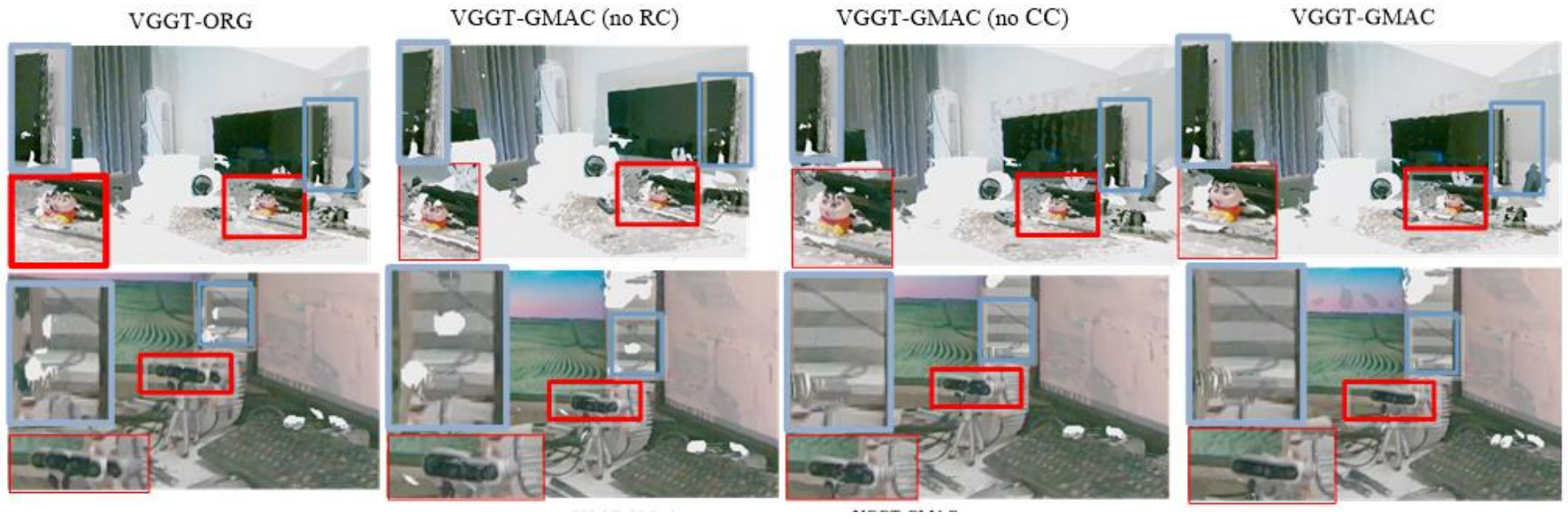

**Fig. 2** Ablation visualization of geometric constraints in GMAC. From left to right: backbone only, without RC, without CC, full GMAC.

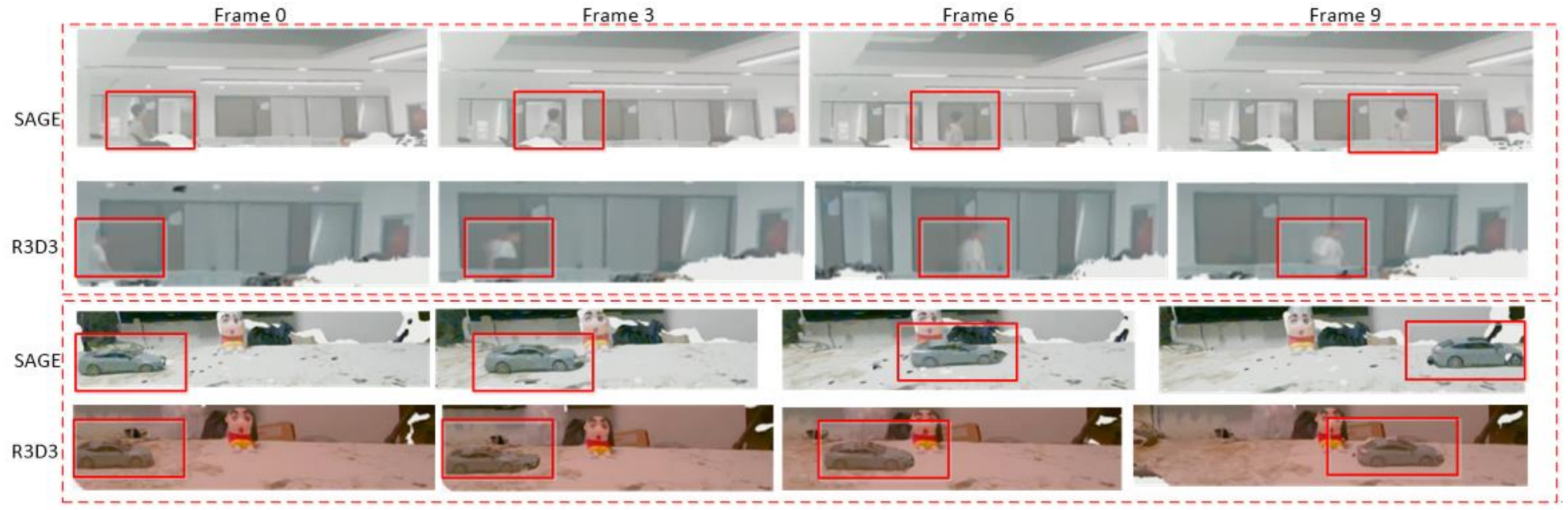

**Fig. 3** Reconstruction effect of SAGE and R3D3 for moving objects with consecutive Multi-Frames in Real-World Scenes. Top: distant view, bottom: close-up

*4) Reliability-Preserving Representation under limited Budget*

Direct evaluation of all pairwise observation relations is costly for large multi-camera rigs. Thus, SAGE adopts fast spatial partition approximation with reliability retention.

All candidate points are first partitioned into non-overlapping spatial regions.

$$\mathcal{P} = \cup_k \mathcal{V}_k \tag{21}$$

For each region cell $\mathcal{V}_k$, the accumulated reliability mass is

$$M_k = \sum_{\boldsymbol{X}_m \in \mathcal{V}_k} r(\boldsymbol{X}_m) \tag{22}$$

The final point cloud is obtained by collecting representative points from all cells under the regional budget constraint.

This approximation preserves reliable geometric observations while naturally suppressing redundant samples in densely populated regions, providing a compact and scalable reconstruction suitable for real-time applications.

*5) Complexity Analysis*

Let $M_{total}$ be the total number of valid depth observations.

Reliability estimation and spatial partitioning both require a single traversal of all observations, resulting in $O(M_{total})$ complexity. In cross-view consistency evaluation, each observation is projected to $|\mathcal{N}_{cam}(i)|$ neighboring cameras, giving $O(M_{total} \cdot |\mathcal{N}_{cam}(i)|)$. Since $|\mathcal{N}_{cam}(i)|$ is bounded, this stage is linear w.r.t. observations.

Within each spatial region, adaptive spatial hashing inserts each candidate observation once, while representative selection only maintains the highest-reliability observation through constant-time comparison, resulting in $O(M_{total})$ complexity.

Therefore, all stages are linear w.r.t. observations under bounded connectivity, with overall complexity $O(M_{total})$. For an RGB-D camera array with $N_{cam}$ cameras and resolution $H \times W$, the overall complexity is $O(N_{cam}HW)$.

In addition, the proposed representation maintains a bounded output size by enforcing $|\mathcal{P}_f| \leq B_{max}$, which prevents the transmitted point set from growing with the number of cameras.

## IV. EXPERIMENTAL RESULTS

Comprehensive tests are carried out on public benchmarks and real-world multi-camera scenes to fully evaluate the proposed framework. Experiments are divided into three parts: standalone evaluation of GMAC, standalone evaluation of SAGE, and joint evaluation of the full SAGE-Flow pipeline.

Unless stated otherwise, all tests run on a desktop workstation with an Intel Core i7-14900KF CPU, 128 GB of RAM, and an NVIDIA RTX 4090 (24 GB) GPU. The camera resolution used is 640×480.

### A. Experimental Setup

*1) Experimental Datasets*

Two types of datasets are used to cover both standard benchmark validation and real-world testing.

1. ScanNet Dataset: It is an indoor RGB-D dataset that

provides accurate camera intrinsics and extrinsic parameters.

2. Real-World Multi-Camera Dataset: Synchronized indoor video sequences are captured with a fixed multi-camera array. Because precise ground-truth extrinsics are hard to obtain in real setups, manually calibrated results are used for evaluation.

*2) Backbone Networks and Baseline Methods*

Experiments are divided into two task categories, and choose appropriate baselines for fair comparison.

1. Extrinsic Estimation Task: a SOTA method for extrinsic parameter prediction (VGGSfM), and two public multi-view reconstruction networks are adopted as backbones for the GMAC module: MapAnything (a SOTA 3D regression model), and VGGT (a representative MRB method). All methods use the same input resolution and inference settings. Only network pruning and structural changes are applied as described in GMAC.

2. Point Cloud Reconstruction Task: R3D3 [21] and VR2Gather [35] (VR2 for short) are used as the primary baselines, as they are among the few public methods supporting synchronized multi-camera real-time reconstruction. Classical RGB-D systems (e.g., KinectFusion, ElasticFusion, and BundleFusion) are excluded because they target single-camera SLAM and are not directly applicable to the MRR setting. 3DGS [23] and HunyuanWorld-Mirror [29] (Hunyuan for short) are included only as upper-bound offline references to characterize the efficiency–quality trade-off.

*3) Evaluation Metrics*

Extrinsic Estimation

1. Rotation Error (degrees, lower is better)

2. Translation Error (millimeters, lower is better)

Point Cloud Reconstruction Quality

1. Plane Fitting Residual (PFR, mm, lower is better): Mean distance of inliers to RANSAC-fitted planes, reflecting geometric smoothness.

2. Photometric NCC Score (PNCC, higher is better): Normalized cross-correlation of $3\times3$ patches between camera pairs for geometry-photometry alignment.

3. Outlier Ratio (OR, %, lower is better): Proportion of outliers after 5 mm radius filtering.

4. Chamfer Distance (CD, mm, lower is better): Evaluating reconstruction accuracy against ground truth.

5. FPS: Real-time inference speed.

6. GPU Memory Usage (GB): GPU memory consumption.

All quantitative tests adopt 100 frames per trial, with all metrics averaged to yield stable evaluation results.

*B. GMAC Module Evaluation*

This section tests GMAC as a standalone extrinsic calibration module. We present quantitative comparisons on ScanNet dataset and visual comparisons on real-world dataset.

*1) Quantitative experimental results*

Quantitative experiments are carried out on the ScanNet dataset. Since the predicted translation of VGGSfM is not guaranteed to preserve the metric scale, only rotation accuracy is reported for direct evaluation. MapAnything and VGGT are further evaluated as MRB backbones before and after GMAC. As shown in Table 1, the raw MRB predictions exhibit noticeable scale inconsistency in translation estimation, leading to inaccurate metric camera poses. After incorporating GMAC, both backbones achieve consistent improvements in rotation and translation accuracy, demonstrating that GMAC effectively leverages implicit geometric priors from multi-view reconstruction networks for metric-scale pose estimation.

TABLE I
EXTRINSIC RECOVERY RESULTS ON SCANNET

| Model Category | Backbone/Method | Rot. Error ↓ | Trans. Error ↓ |
|---|---|---|---|
| Extrinsic Model | VGGSfM (Raw) | 1.31 | N/A |
| MRB | MapAnything (Raw) | 1.19 | 7.35 |
| | MapAnything + GMAC | 0.96 | 3.31 |
| MRB | VGGT (Raw) | 1.04 | 8.61 |
| | VGGT + GMAC | 0.82 | 2.87 |

*3) Ablation Study of GMAC*

To analyze each component's contribution to GMAC, we perform a leave-one-out ablation with VGGT as the backbone on the real-world data.

w/o RC: Remove reprojection consistency.

w/o CC: Remove cycle consistency.

TABLE II
ABLATION RESULTS ON REAL-WORLD

| Method | Rot. Error ↓ | Trans. Error ↓ |
|---|---|---|
| w/o CC | 1.85 | 4.57 |
| w/o RC | 2.12 | 5.51 |
| GMAC | 1.72 | 4.03 |

As shown in Table II, removing RC leads to the largest performance degradation, especially in translation accuracy, demonstrating its critical role in recovering metric scale through depth-guided reprojection alignment. Removing CC also causes performance drops, indicating that cycle consistency helps maintain global geometric coherence and suppress inconsistent pose relations. Combining both constraints enables GMAC to achieve more accurate and stable metric-scale extrinsic estimation. Fig. 2 qualitatively verifies that removing RC results in severe alignment deviation, while removing CC causes slight yet global distortion. The complete GMAC architecture achieves accurate alignment in both local and global domains simultaneously.

*C. SAGE Module Evaluation*

SAGE is evaluated as an independent confidence-driven geometric representation module. To isolate SAGE’s contribution and remove the effect of extrinsic quality, we use ground-truth extrinsics on ScanNet and manually calibrated extrinsics as pseudo-ground-truth on the real-world dataset for all tests. For fair comparison of reconstruction quality, we do not constrain the geometric budget unless otherwise stated.

*1) Static Scene Reconstruction*

The performance of static reconstruction are tested on the ScanNet dataset. Since it is static single-frame reconstruction, FPS indicators are not involved. Results are shown in Table III.

Compared with 3DGS, SAGE leverages direct metric depth measurements and confidence-aware geometric representation achieving comparable geometry quality with significantly lower memory consumption. The lower performance of 3DGS with fewer views is mainly due to insufficient multi-view constraints, while additional views provide more observations and improve its reconstruction quality. Although Hunyuan achieves higher accuracy with learned priors, it requires larger computational overhead, highlighting the advantage of SAGE for real-time multi-view RGB-D reconstruction.

TABLE III
STATIC SCENE RECONSTRUCTION IN SCANNET DATASET

| N | Method | PFR↓ (mm) | PNCC↑ | OR↓ (%) | CD↓ (mm) | GPU Mem (GB) |
|---|---|---|---|---|---|---|
| 4 | Hunyuan | 8.09 | 0.241 | 12.8 | 6.04 | 16.1 |
| | 3DGS | 11.82 | 0.172 | 20.3 | 9.47 | 11.5 |
| | SAGE | 9.57 | 0.197 | 15.5 | 8.02 | 2.2 |
| 8 | Hunyuan | 7.71 | 0.255 | 11.2 | 5.75 | 19.4 |
| | 3DGS | 10.89 | 0.178 | 18.1 | 8.94 | 13.6 |
| | SAGE | 9.13 | 0.208 | 14.2 | 7.60 | 2.6 |

*2) Real-Time Dynamic Scene Reconstruction*

Dynamic scene evaluation is conducted on SAGE, R3D3 and VR2. Results are shown in Table IV. SAGE achieves the best overall performance, while R3D3 suffers from temporal inconsistency accumulation in dynamic environments. VR2 provides efficient streaming but lacks geometry-oriented fusion optimization. By contrast, SAGE's stateless confidence-aware representation maintains reconstruction quality and scalability as camera numbers increase, achieving a better trade-off between accuracy, speed, and memory usage.

TABLE IV
DYNAMIC SCENE RECONSTRUCTION IN REAL-WORLD

| N | Method | PFR↓ (mm) | PNCC↑ | OR↓ (%) | FPS | GPU Mem (GB) |
|---|---|---|---|---|---|---|
| 4 | R3D3 | 13.51 | 0.148 | 19.1 | 8.1 | 8.5 |
| | VR2 | 14.85 | 0.135 | 20.3 | 27.2 | 6.6 |
| | SAGE | 12.24 | 0.172 | 17.6 | 53.2 | 2.7 |
| 8 | R3D3 | 12.86 | 0.156 | 17.9 | 5.8 | 10.1 |
| | VR2 | 14.02 | 0.142 | 18.8 | 22.2 | 7.6 |
| | SAGE | 12.01 | 0.175 | 17.2 | 44.3 | 3.2 |

Fig. 3 further compares dynamic object reconstruction results for medium- and small-scale scenes. R3D3 suffers obvious ghosting, motion trails and surface blurring around moving objects due to temporal consistency-driven error accumulation. SAGE suppresses such artifacts via stateless per-frame confidence-weighted representation, yielding sharp contours and intact surfaces. This visual evidence confirms SAGE's dynamic-scene robustness, matching the results above.

*3) Ablation Study of SAGE*

Ablation experiments are conducted on a 4-camera dynamic scene (real-world data, with ground-truth extrinsics) to validate the effectiveness of each core component of SAGE. Five configurations are designed:

1. w/o CVGC: Remove cross-view geometric consistency and use measurement reliability only for quality estimation.
2. w/o BA: Replace reliability-aware budget allocation with uniform spatial allocation.
3. w/o MR: Remove measurement reliability estimation and assign identical confidence to all observations.
4. w/o RPS: Remove reliability-guided representative point selection and use fixed representative points per cell.

TABLE V
ABLATION RESULTS OF SAGE

| Configuration | PFR↓ (mm) | PNCC↑ | OR↓(%) | FPS |
|---|---|---|---|---|
| w/o CVGC | 14.53 | 0.136 | 18.7 | 61.3 |
| w/o BA | 14.08 | 0.141 | 18.2 | 66.8 |
| w/o MR | 17.11 | 0.121 | 21.6 | 63.9 |
| w/o RPS | 13.41 | 0.146 | 17.8 | 58.2 |
| SAGE | 12.24 | 0.172 | 17.6 | 53.2 |

As shown in Table V, removing any component degrades the performance, validating the effectiveness of each module. MR has the largest impact, indicating the importance of reliable observation estimation. CVGC further improve reconstruction quality through adaptive spatial organization and cross-view consistency, while BA and RPS enhance representation compactness and accuracy.

*4) Geometric Budget Analysis*

Reconstruction performance under different geometric budgets is analyzed. SAGE directly regulates the number of output points through the constraint $\left|P_f\right| \leq B_{max}$. The upper limit of the geometric budget is derived from transmission bandwidth constraints.

For a real 8-camera RGB-D multi-view capture system with a resolution of 640×480, the raw observed point cloud contains roughly 2.5 million points per frame. Based on a transmission bandwidth of 1000 Mbps, a frame rate of 30 FPS, a 10× compression ratio, and 24 bytes per 3D point for storing 3D coordinates and RGB color attributes, the estimated maximum number of points per frame is approximately 1.7 million. Accordingly, we set 1.5 million points per frame as a conservative upper geometric budget and gradually reduce the maximum point threshold to evaluate the corresponding reconstruction performance variations.

The results in Table VI reveal that SAGE sustains stable reconstruction accuracy. Even when the geometric budget is cut to 66% of the upper bound, only minor degradation occurs in reconstruction quality. This finding verifies that SAGE achieves a favorable trade-off between data volume and reconstruction accuracy.

TABLE VI
RECONSTRUCTION QUALITY IN DIFFERENT GEOMETRIC BUDGETS

| Geometric Budget (points/frame) | PFR ↓ (mm) | PNCC ↑ | OR ↓ (%) |
|---|---|---|---|
| 1.5M | 12.31 | 0.170 | 17.9 |
| 1.0M | 13.34 | 0.161 | 18.9 |
| 0.5M | 15.86 | 0.148 | 22.3 |
| 0.2M | 21.12 | 0.119 | 28.5 |

Note: Geometric budget refers to the maximum output point count constrained inside SAGE during reconstruction, not the point count after post-processing compression or transmission coding.

## D. Joint Evaluation: SAGE-Flow Pipeline

This section conducts a systematic evaluation of the integrated SAGE-Flow pipeline from two key aspects: the relationship between reconstruction quality and calibration accuracy, and practical deployment potential including real-time performance and scalability.

### 1) Impact of Extrinsic Accuracy on Reconstruction

With the SAGE module fixed, we evaluate the influence of extrinsic accuracy on reconstruction quality in four camera configration. We apply larger-than-observed translation perturbations to GMAC extrinsics with VGGT as the backbone on ScanNet to evaluate reconstruction robustness.

The perturbed translation is defined as:

$$\hat{\boldsymbol{t}}_i = \boldsymbol{t}_i^{GMAC} + \boldsymbol{\epsilon}_i \tag{23}$$

where $\boldsymbol{\epsilon}_i \in \mathcal{N}(\boldsymbol{0}, \sigma^2 \boldsymbol{I})$ and $\sigma = \rho \left\| \boldsymbol{t}_i^{GMAC} \right\|$, $\rho$ represents different relative translation errors, while rotations remain unchanged.

TABLE VII
RECONSTRUCTION ROBUSTNESS UNDER TRANSLATION PERTURBATIONS OF EXTRINSICS

| Error | PFR↓ (mm) | CD↓(mm) | OR↓(%) |
|---|---|---|---|
| 0% | 9.66 | 8.11 | 15.9 |
| 5% | 10.34 | 8.79 | 17.1 |
| 10% | 11.46 | 9.72 | 18.9 |
| 15% | 13.52 | 11.48 | 21.7 |
| 20% | 15.91 | 13.82 | 26.1 |

The results in Table VII show that reconstruction quality gradually degrades as extrinsic errors increase. This confirms that accurate geometric alignment is important for multi-view fusion. Meanwhile, the degradation remains gradual under moderate perturbations, indicating that SAGE's stateless geometric representation can reduce the impact of locally inconsistent observations.

### 2) Real-Time Performance and Scalability

Real-time performance and hardware scalability are crucial for the practical application of multi-camera reconstruction systems. We therefore test the computational efficiency and scalability of the SAGE-Flow pipeline under different hardware conditions and camera configurations. Specifically, Three representatives hardware platforms are selected to cover different computing power levels:

1. Laptop: i7-9750H / NVIDIA RTX 2060 / 32 GB RAM
2. Desktop: i7-12700 / NVIDIA RTX A2000 / 64 GB RAM
3. Server: i7-14900F / NVIDIA RTX 4090 / 128 GB RAM

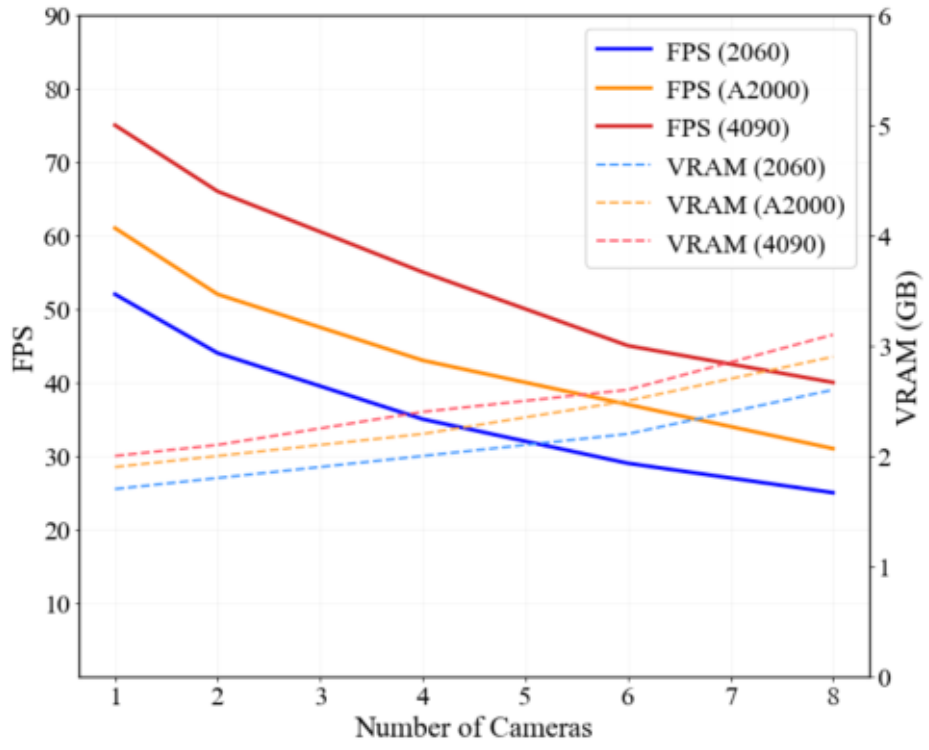

**Fig. 4** FPS of SAGE-FLOW in different configurations.

The results in Fig. 4 show that SAGE-Flow maintains high real-time performance across all platforms; the frame rate decreases approximately linearly with the number of cameras, while VRAM consumption rises linearly with camera count. Even on laptop (minimal compute config), all tested setups achieve real-time frame rates, demonstrating low computational complexity and strong engineering scalability.

# V.CONCLUSION

Real-time multi-camera 3D reconstruction is limited by calibration instability, unreliable fusion, and poor scalability. We propose SAGE-Flow, a decoupled framework separating extrinsic calibration and point cloud reconstruction. GMAC recovers metric camera poses from relative-scale predictions using metric lifting, reprojection, and cycle-consistency constraints without calibration targets or bundle adjustment. SAGE constructs stateless confidence-aware geometric representations by selecting reliable observations under geometric budget constraints, delivering linear complexity and reduced raw data volume for all downstream pipelines.

Experiments on public and real-world systems show improved calibration, reconstruction quality, robustness, and efficiency. GMAC refinement also improves downstream reconstruction, validating the decoupled design.

SAGE-Flow degrades with unreliable matches, sparse overlap, weak textures, motion blur, reflections and missing depth. Future work will boost the algorithm’s robustness, support distributed camera streaming, build a full end-to-end telepresence pipeline, and co-optimize front geometric sampling and rear bitrate allocation.